\documentclass
  {article}
\usepackage{graphicx}
\newcommand{\lora} {\boldmath$\longrightarrow$}
\newcommand{\vecbm}[1]{\mbox{\boldmath#1}}
\newcommand{\vecb}[1]{\mbox{\bf#1}}
\newcommand{\cent}[1] {\begin{center}#1\end{center}}
\newcommand{\lra}  {$\leftrightarrow$}

\newcommand{\mlora} {\bf\longrightarrow}
\newcommand{\doublint}{\int\rule{-3.5mm}{0mm}\int}

\title{ A New Thermodynamics,\\From Nuclei to Stars}

\author{D.H.E. Gross\\
Hahn-Meitner Institute and Freie Universit{\"a}t Berlin,\\
Fachbereich Physik.\\ Glienickerstr. 100\\ 14109 Berlin, Germany}

\begin{document}

\maketitle

\begin{abstract}
  Equilibrium statistics of Hamiltonian systems is correctly described
  by the microcanonical ensemble. Classically this is the manifold of all
  points in the $N-$body phase space with the given total energy.
  Due to Boltzmann's principle, $e^S=tr(\delta(E-H))$, its
  geometrical size is related to the entropy $S(E,N,\cdots)$. This
  definition does not invoke any information theory, no
  thermodynamic limit, no extensivity, and no homogeneity
  assumption, as are needed in conventional (canonical) thermo-statistics.
  Therefore, it describes the equilibrium statistics of extensive
  as well of non-extensive systems. Due to this fact it is the {\em
  fundamental} definition of any classical equilibrium statistics. It can
  address nuclei and astrophysical objects as well. All kind of
  phase transitions can be distinguished sharply and uniquely for even small
  systems. For transitions in nuclear physics the scaling to
  an hypothetical uncharged nuclear matter with an $N/Z-$ ratio like
  realistic nuclei is not needed.
\end{abstract}

\section{Introduction}
Classical Thermodynamics and the theory of phase transitions of
homogeneous and large systems are some of the oldest and best
established theories in physics. It may look strange to add
anything new to it. Let me recapitulate what was told to us since
$>100$ years:
\begin{itemize}
\item Thermodynamics addresses large homogeneous systems at equilibrium
(in the thermodynamic limit
$N\to\infty|_{ N/V=\rho, homogeneous}$).
\item Phase transitions are the positive zeros of the grand-canonical
partition sum $Z(T,\mu,V)$ as function of $e^{\beta\mu}$
(Yang-Lee-singularities). As the partition sum for a finite number
of particles is always positive, zeros can only exist in the
thermodynamic limit $V|_{\beta,\mu}\to\infty$.
\item Micro and canonical ensembles are equivalent.
\footnote{\label{foot} How does one prove this?: The general link
between the microcanonical probability $e^{S(E,N)}$ and the
grand-canonical partition function $Z(T,\mu)$ is by the Laplace
transform:
\begin{eqnarray} Z(T,\mu,V)&=&\doublint_0^{\infty}{\frac{dE}{\epsilon_0}\;
dN\;e^{-[E-\mu N-TS(E,N)]/T}}\\
&=&\frac{V^2}{\epsilon_0}\doublint_0^{\infty} {de\;dn\;e^{-V[e-\mu
n-Ts(e,n)]/T}}\label{grandsum}\\ &\approx&\hspace{2cm}e^{\mbox{
const.+lin.+quadr.}}\label{saddlepoint}
\end{eqnarray}\\~\\
If $s(e,n)$ is concave then there is a single point $e_s$,$n_s$
with
\begin{eqnarray}
\frac{1}{T}&=&\left.\frac{\partial S}{\partial
E}\right|_s\nonumber\\ \frac{\mu}{T}&=&-\left.\frac{\partial
S}{\partial N}\right|_s\nonumber
\end{eqnarray}
In the thermodynamic limit ($V\to\infty$) the quadratic
approximation in equ.(\ref{saddlepoint}) becomes exact and there
is a one to one mapping of the microscopic mechanical
$e=E/V,n=N/V$ to the intensive $T,\mu$. {\bf This on the other
hand shows explicitely that the intensive variables $T,\mu$ are
ill defined in small systems like nuclei}. They, as well as free
energy $F=E-TS$, or enthalpy $H=E+PV$, etc.., should not be used.}
\item Thermodynamics works with intensive variables $T,P,\mu$.
\item Unique Legendre mapping $T\to E$.
\item Heat only flows from hot to cold (Clausius)
\item Second Law only in infinite systems when the Poincarr\'{e}
recurrence time becomes infinite (much larger than the age of the
universe (Boltzmann)).
\end{itemize}

 Under these constraint only a tiny part of the real world of
equilibrium systems can be treated. The ubiquitous non-homogeneous
systems: nuclei, clusters, polymer, soft matter (biological)
systems, but also the largest, astrophysical systems are not
covered. Even normal systems with short-range coupling at phase
separations are inhomogeneous and can only be treated within
conventional homogeneous thermodynamics (e.g. van-der-Waals
theory) by bridging the unstable region of negative
compressibility by an artificial Maxwell construction. Thus even
the original goal, for which Thermodynamics was invented some
$150$ years ago, the description of steam engines is only
artificially solved. There is no (grand-)canonical ensemble of
phase separated and, consequently, inhomogeneous, configurations.
It has a deep reason as I will discuss below: here the systems
have a {\em negative} heat capacity (resp. susceptibility). This,
however, is impossible in the (grand-)canonical theory where
$c\propto (\delta E)^2$

\subsection{Example: Why one should not use conventional scaling
theory to fix phase transitions in Nuclear Physics}

Still many authors addressing phase-transitions in nuclear physics
believe they should link these phenomena by finite size scaling to
"nuclear matter" in the thermodynamic limit. This is the way how
phase transitions can be located in conventional canonical
statistics of ordinary extensive systems with short-range
interactions.

However, infinite nuclear matter with the same $N/Z$ ratio as
ordinary nuclei does not exist. The Coulomb self-energy of such a
system  would be infinite. Therefore the Coulomb interaction
between the fragments of a hot nucleus must be switched off in
such studies. There is no sense in using conventional scaling
analysis to relate transition phenomena seen in nuclear collisions
to anything that does not exist in real life. The Coulomb field
has an important effect on the fragmentation of real nuclei (see
e.g. the article by LeFevre at this conference) which one must
understand by statistical fragmentation simulations. Thus, in the
conventional approach there is the only choice, either one
describes realistic but small nuclei without any connection to
phase transitions. Or one has a scaling theory to something that
does not exist in reality: uncharged nuclear matter with fixed
neutron to proton ratio.

If anything like phase-transitions exist in real nuclei then only
in the sense of a new generalized thermodynamics. In the rest of
this paper I will describe such a generalization which takes
Boltzmann's principle serious and avoids the thermodynamic limit.
This opens Thermodynamics to the much larger world of
non-extensive systems. The most prominent example are of course
self-gravitating astrophysical systems which I will discuss here
also.
\section{Boltzmann's principle}
   The Microcanonical ensemble is the ensemble (manifold)
   of all possible points in the $6N$ dimensional phase space at
   the prescribed sharp energy $E$:
\begin{eqnarray*}
W(E,N,V)&=&\epsilon_0 tr\delta(E-H_N)\\
tr\delta(E-H_N)&=&\int{\frac{d^{3N}p\;d^{3N}q}{N!(2\pi\hbar)^{3N}}
\delta(E-H_N)}.
\end{eqnarray*}
Thermodynamics addresses the whole ensemble. It is ruled by
  the topology of the geometrical size $W(E,N,\cdots)$,
 Boltzmann's principle:
 \begin{equation}
\fbox{\fbox{\vecbm{S=k*lnW}}}
\end{equation}
is the most fundamental definition of the entropy $S$.  Entropy
and with it micro-canonical thermodynamics has therefore a pure
mechanical, geometrical foundation. No information theoretical
formulation is needed. Moreover, in contrast to the canonical
entropy, $S(E,N,..)$ is everywhere single valued and multiple
differentiable. There is no need for extensivity, no need of
concavity, no need of additivity, and no need of the thermodynamic
limit. This is a great advantage of the geometric foundation of
equilibrium statistics over the conventional definition by the
Boltzmann-Gibbs canonical theory. However, addressing entropy to
finite eventually small systems we will face a new problem with
Zermelo's objection against the monotonic rise of entropy, the
Second Law, when the system approaches its equilibrium. This is
discussed elsewhere \cite{gross183,gross192}. A further comment:
In contrast to many authors like Schr\"odinger
\cite{schroedinger46} our ensemble is {\em not} an ensemble of
non-interacting replicas of the considered system which may
exchange energy. I do {\em not} consider the different ways to
distribute energy over the different replicas. I consider the
manifold of the same system at the precisely given energy under
all possible different distributions of the momenta and positions
of its constituents (particles) in the $6N$-dimensional phase
space. The result is then the average behaviour when one does not
know the precise position and momentum of every particle but only
the total energy.
\section{Topological properties of $S(E,\cdots)$}

The topology of the Hessian of $s(E,\cdots)$, the determinant of
curvatures of $s(e,n)$, determines completely all kinds phase
transitions. This is evidently so, because $e^{S(E)-E/T}$ is the
weight of each energy in the canonical partition sum at given $T$,
see footnote \ref{foot}. Consequently, at phase separation this
has at least two maxima, the two phases. And in between two maxima
there must be a minimum where the curvature of $S(E)$ is positive.
I.e. the positive curvature detects phase separation. This is of
course also in the case of several conserved control parameters.
\begin{eqnarray}
d(e,n)&=&\left\|\begin{array}{cc} \frac{\partial^2 s}{\partial
e^2}& \frac{\partial^2 s}{\partial n\partial e}\\ \frac{\partial^2
s}{\partial e\partial n}& \frac{\partial^2 s}{\partial n^2}
\end{array}\right\|=\lambda_1\lambda_2 \label{curvdet}\\
\lambda_1&\ge&\lambda_2\hspace{1cm}\mbox{\lora eigenvectors
:}\hspace{1cm} {\boldmath\vecbm{$v$}_1,\vecbm{$v$}_2}\nonumber
\end{eqnarray}
Of course for a finite system each of these maxima of
$S(E,\cdots)-E/T$ have a non-vanishing width. These are intrinsic
fluctuations in each phase.
\subsection{ Unambiguous signal of phase transitions in a "Small"
system} 
\begin{figure}
\includegraphics*[bb =0 0 290 180, angle=0, width=12cm,
clip=true]{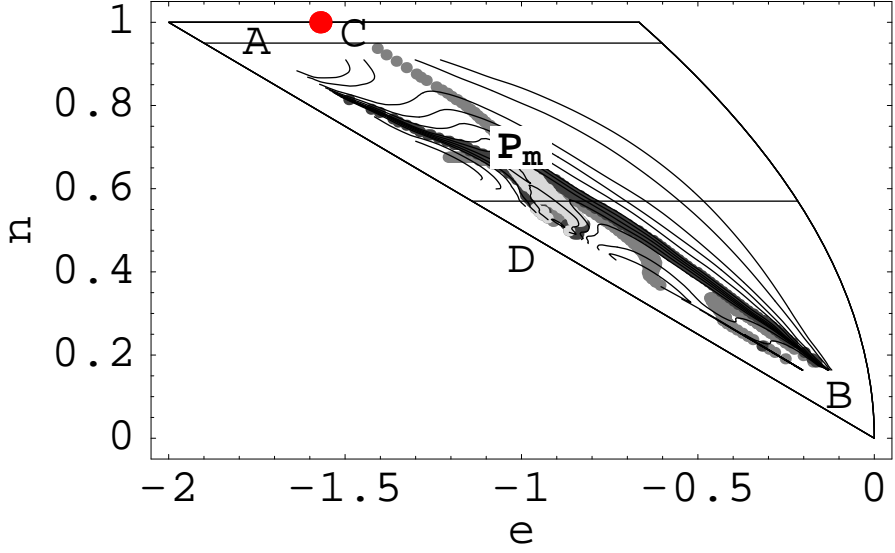}
\caption{Global phase diagram or conture plot of the curvature
determinant
  (Hessian), eqn.~(\ref{curvdet}), of the 2-dim Potts-3 lattice gas
  with $50*50$ lattice points, $n$ is the number of particles per
  lattice point, $e$ is the total energy per lattice point. The line
   (-2,1) to (0,0) is the ground-state energy of the lattice-gas
   as function of $n$. The most right curve is the locus of configurations
    with completely random spin-orientations (maximum entropy). The whole
    physics of the model plays between these two boundaries.  At the
    dark-gray lines the Hessian is $\det=0$,this is the boundary of the
    region of phase separation (the triangle $AP_mB$) with a negative
    Hessian ($\lambda_1>0,\lambda_2<0$).  Here, we have
    Pseudo-Riemannian geometry. At the light-gray lines
    is a minimum of $\det(e,n)$ in the direction of the largest
    curvature (\vecbm{v}$_{\lambda_{max}}\cdot$\vecbm{$\nabla$}$\det=0$) and
    $\det=0$,these are lines of second order transition. In the triangle
    $AP_mC$ is the pure ordered (solid) phase ($\det>0, \lambda_1<0$).
    Above and right of the line $CP_mB$ is the pure disordered (gas) phase
    ($\det>0, \lambda_1<0$). The crossing $P_m$ of the boundary lines is a
    multi-critical point. It is also the critical end-point of the region
    of phase separation ($\det<0,\lambda_1>0,\lambda_2<0$).  The light-gray
    region around the multi-critical point $P_m$ corresponds to a flat,
    horizontal region of $\det(e,n)\sim 0$ and {\boldmath
    \vecbm{$\nabla$}\mbox{$\lambda_1$}{\boldmath$\sim 0$}} and consequently
    to a somewhat extended cylindrical region of $s(e,n)$,
    details see \protect\cite{gross173,gross174}; $C$ is the analytically
   known position of the critical point (second order transition) which the
   ordinary $q=3$ Potts model (without vacancies){\em would have in the
   thermodynamic limit}.}\label{det}
\end{figure}

Nevertheless, the whole zoo of phase-transitions can be
sharply seen and distinguished. This is here demonstrated for the
Potts-gas model on a two dimensional lattic of {\em finite} size
of $50\times 50$ lattice points, c.f. fig.(\ref{det}).
\newpage
\subsection{Systematic of phase transitions in the micro-canonical ensemble without
invoking the thermodynamic limit}

\begin{itemize}
\item A single stable phase of course with some intrinsic fluctuations (width) by
$\lambda_1<0$. Here $s(e,n)$ is concave (downwards bending) in
both directions.  Then there is a one to one mapping of the
canonical \lra the micro-ensemble. \cent{\includegraphics*[bb =
53 14 430 622, angle=-90, width=9cm, clip=true]{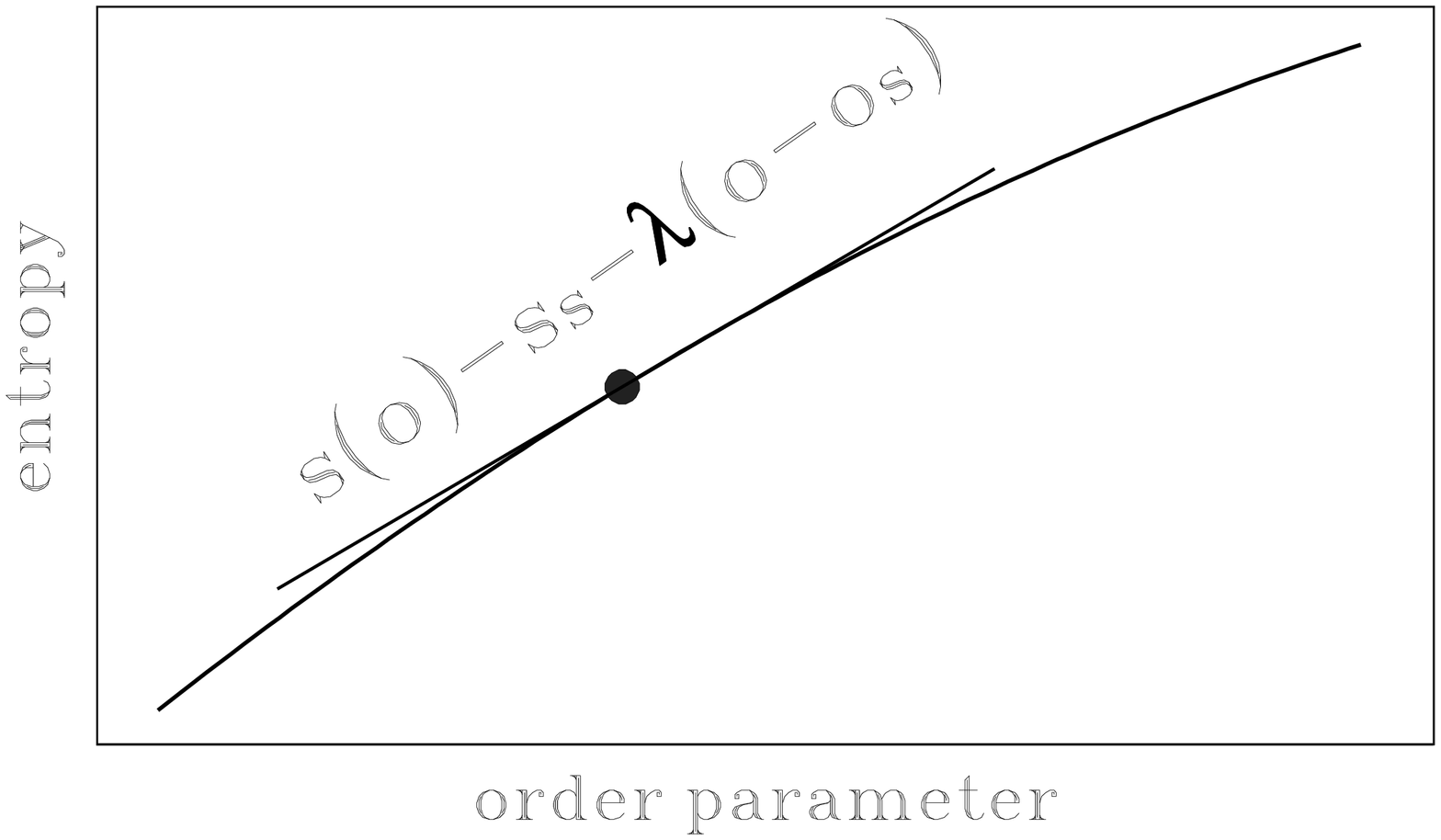}}
\item A transition of first order with phase separation
  and surface tension is indicated by
$\lambda_1(e,n)>0$. $s(e,n)$ has a convex intruder (upwards
bending) in the direction
  $\vecb{v}_1$ of the largest curvature $\ge 0$  which can be identified with the
  order parameter \cite{gross174}. Three solutions of
\begin{eqnarray}
\beta=\frac{1}{T}&=&\left.\frac{\partial S}{\partial
E}\right|_s\nonumber\\ \nu=\frac{\mu}{T}&=&-\left.\frac{\partial
S}{\partial N}\right|_s\nonumber
\end{eqnarray} determine the intensive temperature $T=1/\beta$ and
the chemical potential $T\nu$.
\cent{\includegraphics*[bb = 105 15 463 623, angle=-90, width=9cm,
  clip=true]{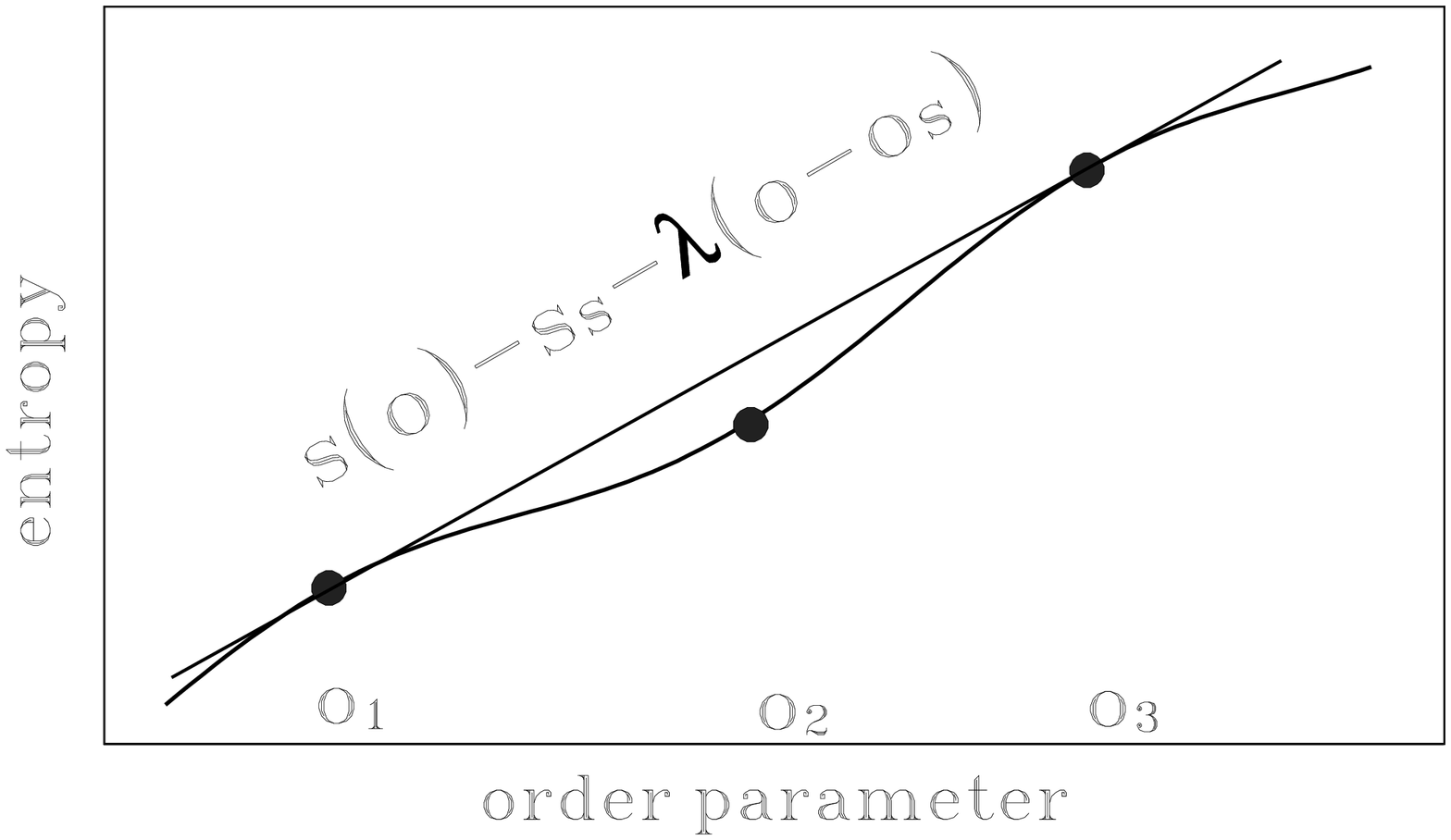}}
In the thermodynamic limit the whole region \{$o_1,o_3$\} is
mapped into a single point in the canonical ensemble which is
consequently non-local in $o$. I.e. if the curvature of $S(E,N)$
is $\lambda_1\ge 0$ both ensembles are not equivalent even in the
limit.
\item A continuous (``second order'') transition
with vanishing surface tension, where two neighboring phases
become indistinguishable. This is indicated by a line with
$\lambda_1=0$ and extremum of $\lambda_1$ in the direction of
order parameter $
\vecb{v}_{\lambda=0}\cdot\vecb{$\nabla$}\lambda_1=0$. These are
the catastrophes of the Laplace transform $E\to T$.
\end{itemize}
\subsection{CURVATURE}

We saw that the curvature (Hessian) of $S(E,N,\cdots)$ controls
the phase transitions. What is the physics behind the curvature?
For short-range force it is linked to the interphase surface
tension.

\begin{table}[h]
\caption{Parameters of the liquid--gas transition of small
  sodium clusters (MMMC-calculation~\protect\cite{gross174}) in
  comparison with the bulk for a rising number $N_0$ of atoms,
  $N_{surf}$ is the average number of surface atoms (estimated here as
  $\sum{N_{cluster}^{2/3}}$) of all clusters with $N_i\geq2$ together.
  $\sigma/T_{tr}=\Delta s_{surf}*N_0/N_{surf}$ corresponds to the
  surface tension. Its bulk value is adjusted
  to agree with the experimental values of the $a_s$ parameter which
  we used in the liquid-drop formula for the binding energies of small
  clusters, c.f.  Brechignac et al.~\protect\cite{brechignac95}, and
  which are used in this calculation~\cite{gross174} for the individual
  clusters.}
\begin{center}
\renewcommand{\arraystretch}{1.4}
\setlength\tabcolsep{5pt}
\begin{tabular} {|c|c|c|c|c|c|} \hline
&$N_0$&$200$&$1000$&$3000$&\vecb{bulk}\\ 
\hline \hline &$T_{tr} \;[K]$&$940$&$990$&$1095$&\vecb{1156}\\
\cline{2-6} &$q_{lat} \;[eV]$&$0.82$&$0.91$&$0.94$&\vecb{0.923}\\
\cline{2-6} {\bf Na}&$s_{boil}$&$10.1$&$10.7$&$9.9$&\vecb{9.267}\\
\cline{2-6} &$\Delta s_{surf}$&$0.55$&$0.56$&$0.44$&\\ \cline{2-6}
&$N_{surf}$&$39.94$&$98.53$&$186.6$&\vecbm{$\infty$}\\ \cline{2-6}
&$\sigma/T_{tr}$&$2.75$&$5.68$&$7.07$&\vecb{7.41}\\ \hline
\end{tabular}
\end{center}
\end{table}
\begin{figure}[h]\cent{
\includegraphics*[bb = 99 57 400 286, angle=-0, width=9cm,
clip=true]{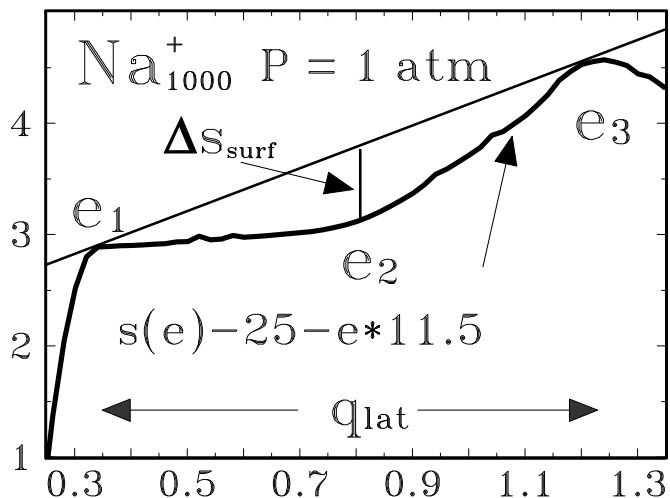}} \caption{MMMC~\protect\cite{gross174}
simulation of the entropy $s(e)$ per atom ($e$ in eV per atom) of
a system of
  $N_0=1000$ sodium atoms at an external pressure of 1 atm.  At the
  energy $e\leq e_1$ the system is in the pure liquid phase and at
  $e\geq e_3$ in the pure gas phase, of course with fluctuations. The
  latent heat per atom is $q_{lat}=e_3-e_1$.  \underline{Attention:}
  the curve $s(e)$ is artifically sheared by subtracting a linear
  function $25+e*11.5$ in order to make the convex intruder visible.
  {\em $s(e)$ is always a steeply monotonic rising function}.  We
  clearly see the global concave (downwards bending) nature of $s(e)$
  and its convex intruder. Its depth is the entropy
  loss due to additional correlations by the interfaces. It scales $\propto
  N^{-1/3}$. From this one can calculate the surface
  tension per surface atom
  $\sigma_{surf}/T_{tr}=\Delta s_{surf}*N_0/N_{surf}$.  The double
  tangent (Gibbs construction) is the concave hull of $s(e)$. Its
  derivative gives the Maxwell line in the caloric curve $T(e)$ at
  $T_{tr}$. In the thermodynamic limit the intruder would disappear and $s(e)$
  would approach the double tangent from below.  Nevertheless, even
  there, the probability $\propto e^{Ns}$ of configurations with
  phase-separations are suppressed by the
  (infinitesimal small) factor $e^{-N^{2/3}}$ relative to the pure
  phases and the distribution remains {\em strictly bimodal in the
    canonical ensemble}. The region $e_1<e<e_3$ of phase separation
  gets lost.\label{naprl0f}}
\end{figure}
\clearpage
\subsection{Heat can flow from cold to hot}
\begin{figure}[h]
\begin{center}
\includegraphics*[bb =38 8 387 611, angle=-180, width=6.7 cm,
clip=true]{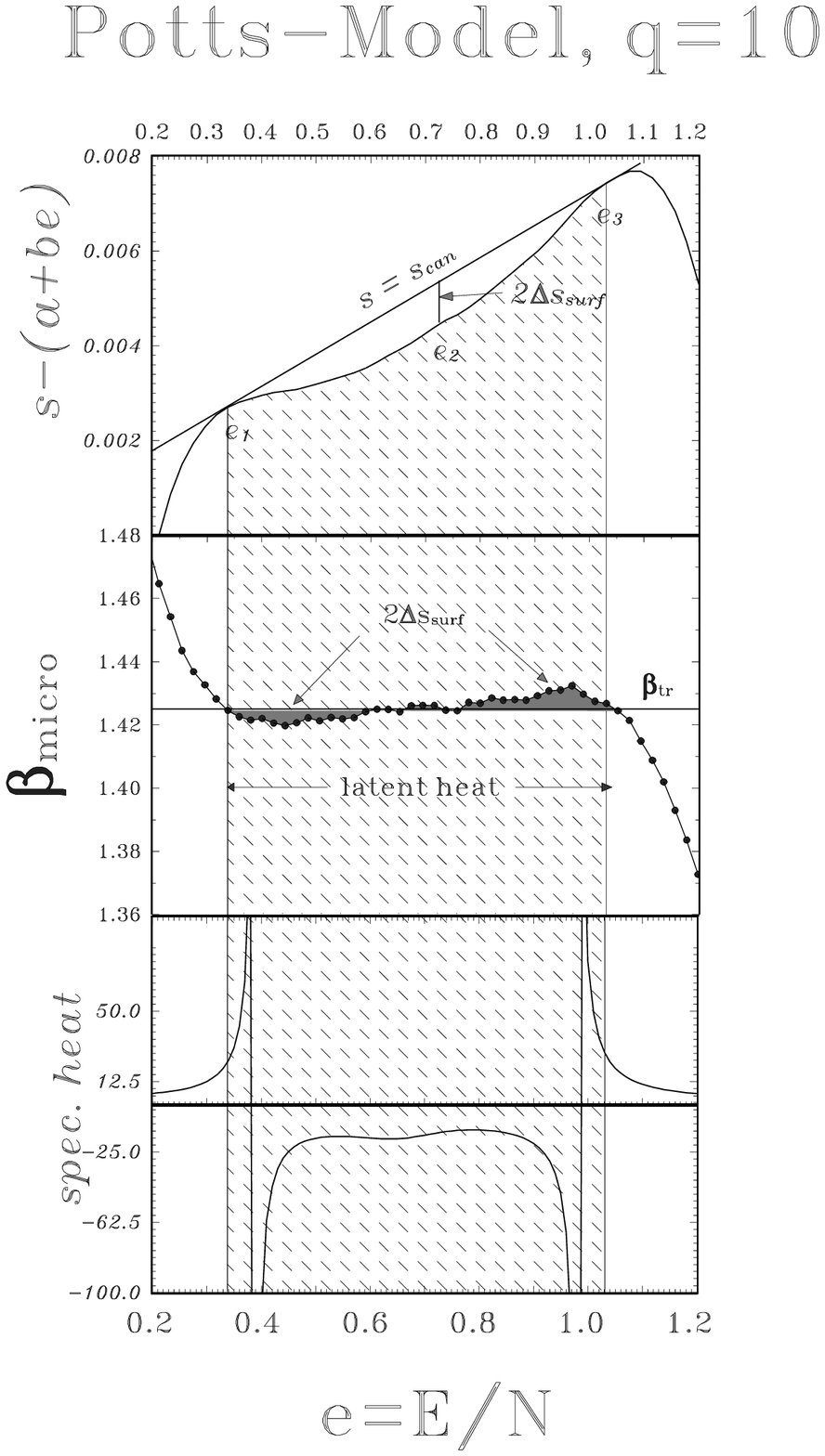}\caption{ Potts model, ($q=10$) in the region
of phase separation. At $e_1$ the system is in the pure ordered
phase, at $e_3$ in the pure disordered phase. A little above $e_1$
the temperature $T=1/\beta$ is higher than a little below $e_3$.
Combining two parts of the system: one at the energy $e_1+\delta
e$ and at the temperature $T_1$, the other at the energy
$e_3-\delta e$ and at the temperature $T_3<T_1$ will equilibrize
with a rise of its entropy, a drop of $T_1$ (cooling) and an
energy flow (heat) from $3\to 1$: i.e.: Heat flows from cold to
hot! Clausius formulation of the Second Law is violated.
Evidently, this is not any peculiarity of gravitating systems!
This is a generic situation within classical thermodynamics even
of systems with short-range coupling and {\em has nothing to do
with long range interaction.}}\vspace{-2cm}
\end{center}
\end{figure}
\clearpage
\section{ Negative heat capacity as signal for a phase
transition of first order.}
\subsection{Nuclear Physics}
A very detailed illustration of the appearance of negative heat
capacities is given by d.Agostino et al. \cite{dAgostino00}. Here
I want to remember one of the oldest experimental finding of a
"back"-bending caloric curve in Nuclear Physics.
\begin{figure}[h]
\includegraphics*[bb = 8 29 495 597, angle=-90, width=10 cm,  
clip=true ]{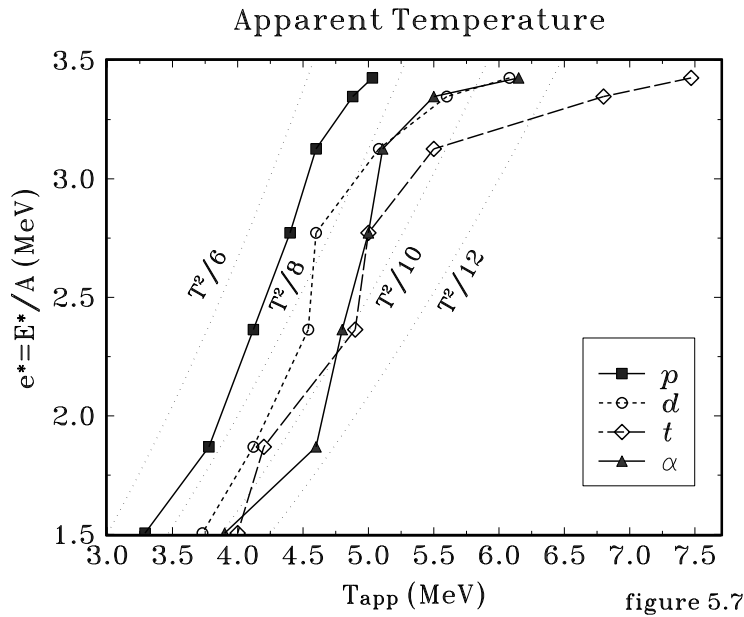}                                         
\caption{Experimental excitation energy per nucleon $e^*$ versus
  apparent temperature $T_{app}$ for backward $p$,
  $d$, $t$ and $\alpha$ together with heavy evaporation residues out of
  incomplete fusion of  $701$ Mev $^{28}$Si$+^{100}$Mo. The dotted curves give the
  Fermi-gas caloric curves for the level-density
  parameter a = 6 to 12. (Chbihi et al. Eur.Phys.J. A 1999)}
\end{figure}
\clearpage
\subsection{Atomic clusters}
Here I show the simulation of a typical fragmentation transition
of a system of $3000$ sodium atoms interacting by realistic
(many-body) forces. To compare with usual macroscopic conditions,
the calculations were done at each energy using a volume $V(E)$
such that the microcanonical pressure $P=\frac{\partial
S}{\partial V}/\frac{\partial S}{\partial E}=1$atm. The inserts
above give the mass distribution at the various points. The label
"4:1.295" means 1.295 quadrimers on average.  This gives a
detailed insight into what happens with rising excitation energy
over the transition region: At the beginning ($e^*\sim 0.442$ eV)
the liquid sodium drop evaporates 329 single atoms and 7.876
dimers and 1.295 quadrimers on average. At energies $e>\sim 1$eV
the drop starts to fragment into several small droplets
("intermediate mass fragments") e.g. at point 3: 2726 monomers,80
dimers,$\sim$5 trimers, $\sim$15 quadrimers and a few heavier ones
up to 10-mers. The evaporation residue disappears. This
multifragmentation finishes at point 4. It induces the strong
backward swing of the caloric curve $T(E)$. Above point 4 one has
a gas of free monomers and at the beginning a few dimers. This
transition scenario has a lot similarity with nuclear
multifragmentation. It is also shown how the total interphase
surface area, proportional to $N_{eff}^{2/3}=\sum_i N_i^{2/3}$
with $N_i\ge 2$ ($N_i$ the number of atoms in the $i$th cluster)
stays roughly constant up to point 3 even though the number of
fragments ($N_{fr}=\sum_i$) is monotonic rising with increasing
excitation.
\begin{figure}[h]\vspace*{-0.5
cm}
 \cent{\includegraphics [bb = 69 382 545
766, angle=-0, width=12cm, clip=true]{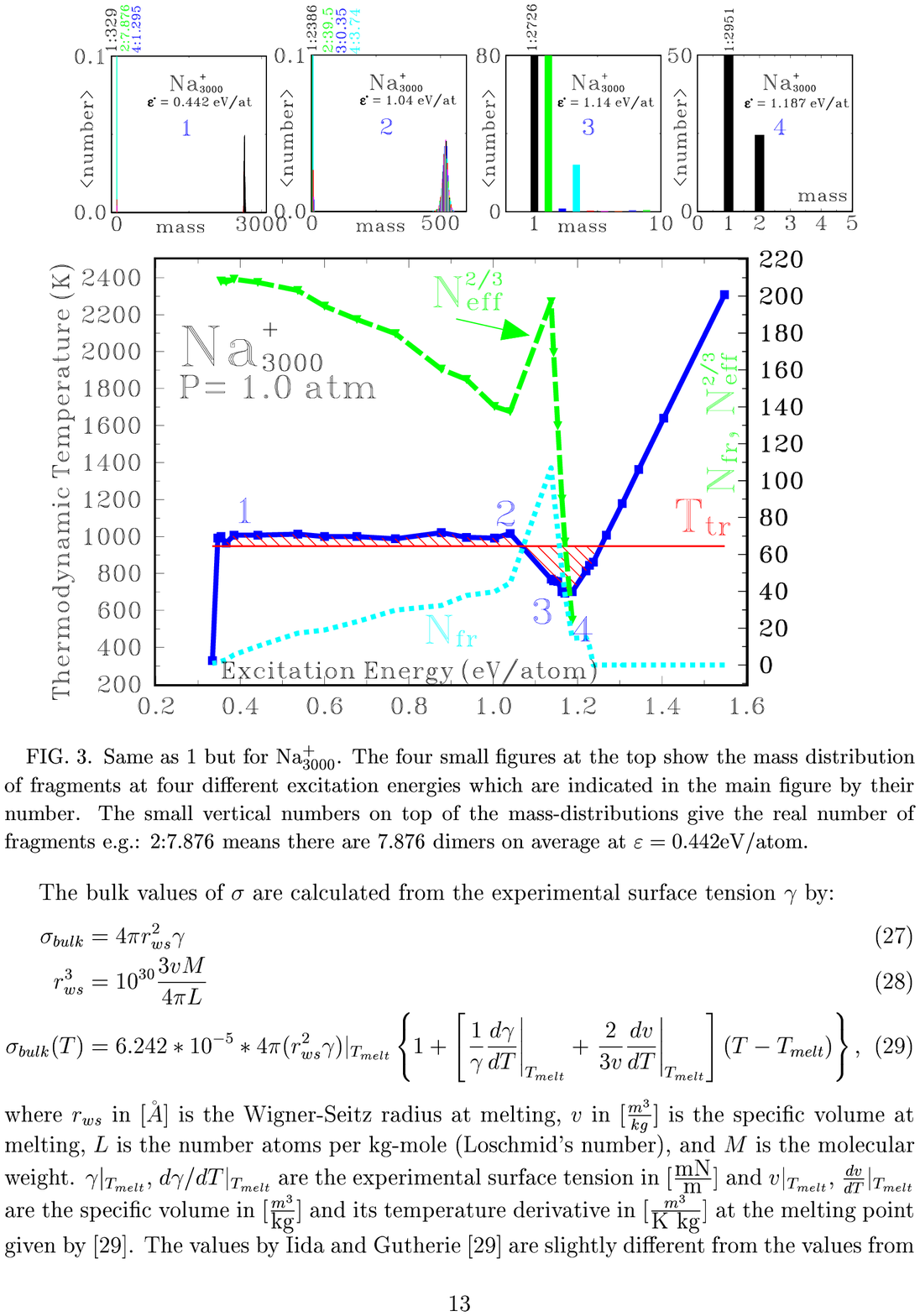}} \vspace*{-0.5
cm}\caption{Cluster fragmentation}
\end{figure}
\clearpage

\subsection{Stars}
    Self-gravitation leads to a non-extensive potential energy $\propto
N^2$.  No thermodynamic limit exists for $E/N$ and no canonical
treatment makes sense. At negative total energies these systems
have a negative heat capacity.  This was for a long time
considered as an absurd situation within canonical statistical
mechanics with its thermodynamic ``limit''. However, within our
geometric theory this is just a simple example of the
pseudo-Riemannian topology of the microcanonical entropy $S(E,N)$
provided that high densities with their non-gravitational physics,
like nuclear hydrogen burning, are excluded. We treated the
various phases of a self-gravitating cloud of particles as
function of the total energy and angular momentum, c.f. the quoted
paper. Clearly these are the most important constraint in
astrophysics.
    \begin{figure}[h]
    \includegraphics*[bb =88 401 522 630, angle=-0, width=12 cm,
    clip=true]{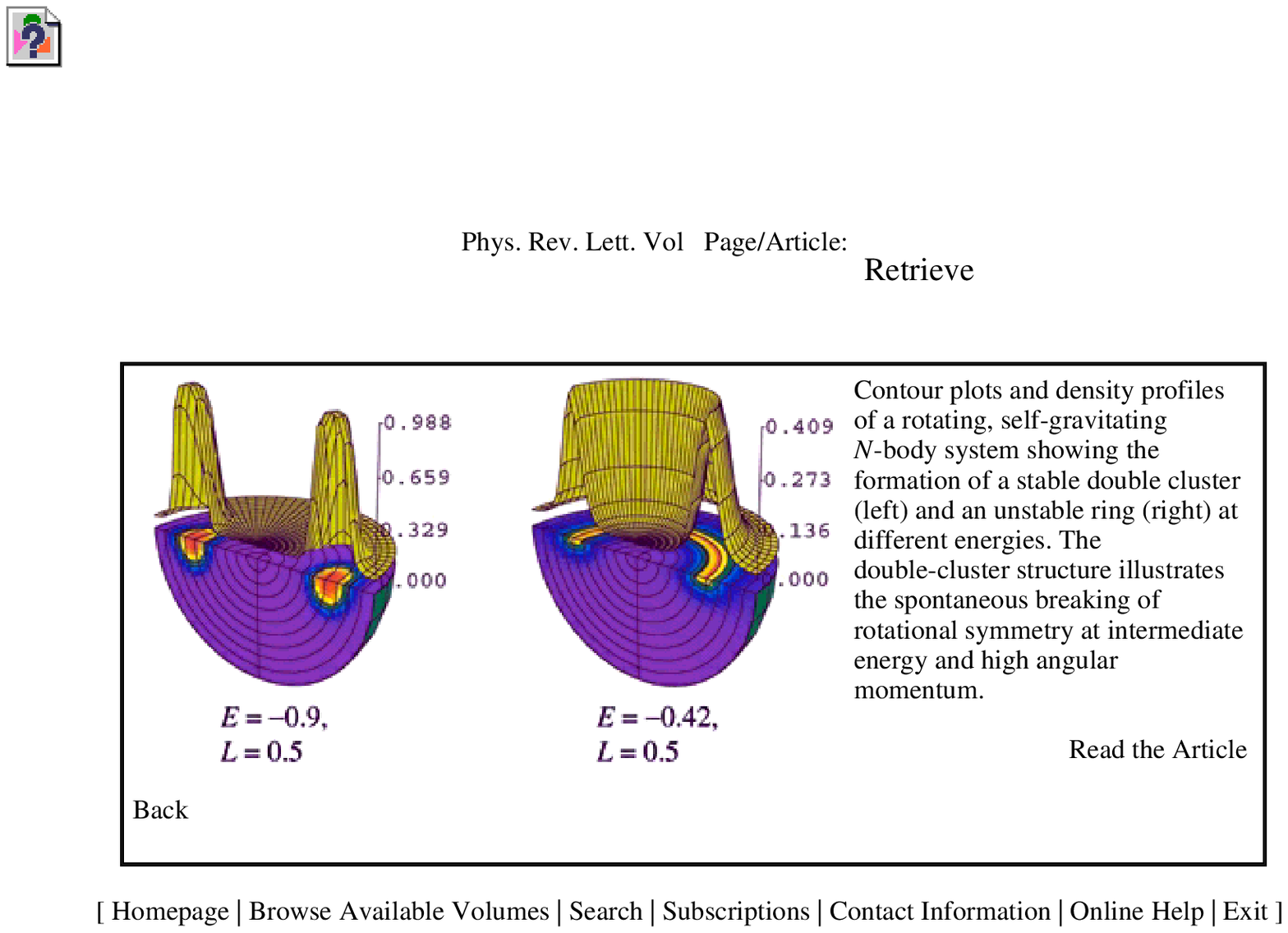}
    \caption{Phases and Phase-Separation
    in Rotating, Self-Gravitating Systems,
    Physical Review Letters--July 15, 2002, cover-page, by
    (Votyakov, Hidmi, De Martino, Gross}
    \end{figure}
\begin{figure}
\includegraphics[bb =72 54 533 690,width=8cm,angle=-90,clip=true]{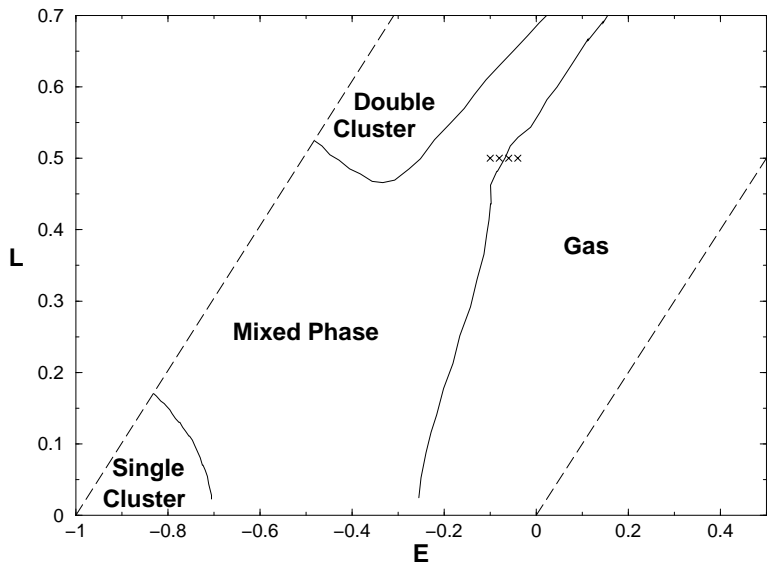}
\caption{Microcanonical phase-diagram of a cloud of
self-gravitating and rotating system as function of the energy and
angular-momentum. Outside the dashed boundaries only some singular
points were calculated. In the mixed phase the largest curvature
$\lambda_1$ of $S(E,L)$ is positive. Consequently the heat
capacity or the correspondent susceptibility is negative. This is
of course well known in astrophysics. However, the new and
important point of our finding is that within microcanonical
thermodynamics this is {\em a generic property of all phase
transitions of first order, whether there is a short- or a
long-range force that organizes the system}.}
\end{figure}

\clearpage
\section{Conclusion}

Entropy has a simple and elementary definition by the {\em area}
$e^{S(E,N,\cdots)}$ of the microcanonical ensemble in the $6N$
dim. phase space. Canonical ensembles are not equivalent to the
micro-ensemble in the most interesting situations:
\begin{enumerate}
\item at phase-separation (\lora heat engines !), one gets
 inhomgeneities, and a negative heat capacity or some other negative susceptibility,
\item  Heat can flow from cold to hot.
\item phase transitions can be localized sharply and unambiguously in
small classical or quantum systems, there is no need for finite
size scaling to identify the transition.
\item also really large self-gravitating systems can be addressed.
\end{enumerate}
Entropy rises during the approach to equilibrium, $\Delta S\ge 0$,
also for small mixing systems. i.e. the Second Law is valid even
for small systems \cite{gross183,gross192}.

With this geometric foundation thermo-statistics applies not only
to extensive systems but also to non-extensive ones which have no
thermodynamic limit. For the application to Nuclear Physics I
believe one should not define phase transitions by scaling
arguments towards a non-existing "nuclear matter". This may
overlook the non-extensivity of realistic nuclear systems. The
Coulomb field of a fragmenting nucleus has an important influence
on the mass and charge distribution of multifragmentation.


\end{document}